\documentclass[aps,pra,twocolumn,floatfix]{revtex4}
\usepackage{amsmath}
\usepackage{amssymb}
\usepackage{bm}
\usepackage{epsfig}

\def\be{\begin{equation}}
\def\ee{\end{equation}}
\def\bea{\begin{eqnarray}}
\def\eea{\end{eqnarray}}

\begin{document}

\title{Dynamical evolution of correlated spontaneous emission of a single
photon from a uniformly excited cloud of N atoms}
\author{Anatoly A. Svidzinsky, Jun-Tao Chang and Marlan O. Scully}
\affiliation{{\small Institute for Quantum Studies and Department of Physics, Texas A\&M
University, College Station, Texas 77843 }\\
Applied Physics and Materials Science Group, Engineering Quad, Princeton
University, Princeton, New Jersey 08544 }
\date{\today }

\begin{abstract}
We study the correlated spontaneous emission from a dense spherical cloud of 
$N$ atoms uniformly excited by absorption of a single photon. 
We find that the decay of such a state depends on the relation between an
effective Rabi frequency $\Omega \propto \sqrt{N}$ and the time of photon
flight through the cloud $R/c$. If $\Omega R/c<1$ the state exponentially
decays with rate $\Omega ^{2}R/c$ and the state life time is greater then $%
R/c$. In the opposite limit $\Omega R/c\gg 1$, the coupled atom-radiation
system oscillates 
between the collective Dicke state (with no-photons) and the atomic ground
state (with one photon) with frequency $\Omega $ while decaying at a rate $%
c/R$. 
\end{abstract}

\maketitle

\begin{widetext}

\begin{figure}[h]

\epsfxsize=0.3\textwidth\epsfysize=0.25\textwidth
\center{\includegraphics[width=17cm]{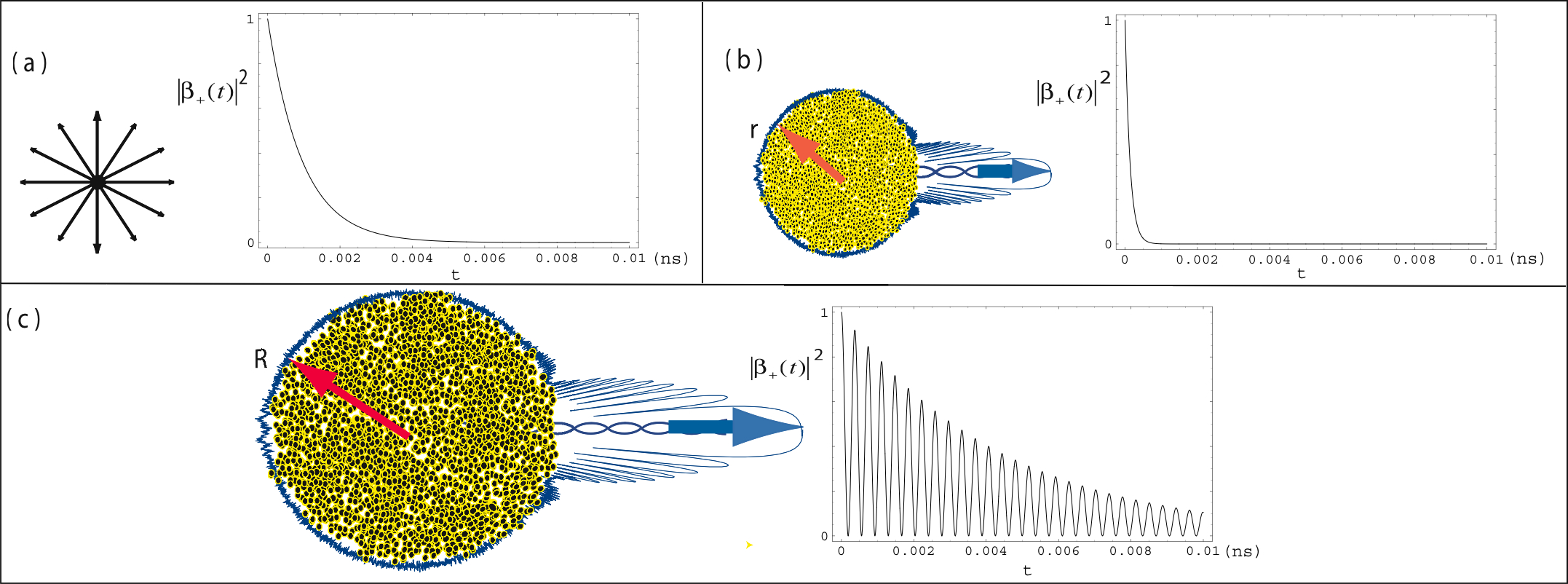}} 
\caption{
Comparison of the different dynamical behavior for the correlated
spontaneous emission from an $N$ atomic cloud described by the $|+\rangle _{{%
\mathbf{k}}_{0}}$ state of Fig. 2b. The single atom spontaneous decay 
life time is taken to be $\tau _{0}=10$ ns ($\gamma =1/2\tau _{0}$). We
assume that atomic density is $10^{16}$cm$^{-3}$ and the resonant photon
wavelength is $\lambda =1\mu $m. Plot (a) corresponds to the case when cloud
radius is equal to $\lambda /2$, hence the number of atoms is $N_{a}=5240$,
then the state decay time is 
$\tau _{a}=\tau _{0}/N_{a}=1.9\times 10^{-3}$ ns. In plot (b) the cloud
radius is $r=10\lambda $, 
$N_{b}=4.2\times 10^{7}$, $\tau _{b}=32\pi ^{2}r^{2}\tau _{0}/27N_{b}\lambda
^{2}=2.7\times 10^{-4}$ ns. In plot (c) the radius of the atomic cloud is $%
R=1$ mm which yields 
$N_{c}=4.1\times 10^{13}$, $\tau _{c}=8R/6c=4.4\times 10^{-3}$ ns, while the
period of oscillations is $2\pi /\Omega =0.74\times 10^{-3}$ ns.
}
\label{Fig1}
\end{figure}

\end{widetext}

The problem of a single photon absorbed by a cloud of $N$ atoms followed by
correlated spontaneous emission is a problem of long standing interest.
Dicke \cite{Dick54} first noted that the radiation rate from a small dense
cloud is abnormal. In his words \textquotedblleft ... the greatest radiation
intensity anomaly occurs in the transition to the ground state" \cite%
{Dicke64}. In particular he showed that the collective decay rate for the
symmetric state with one excitation is $\Gamma _{N}=N\gamma $ (as indicated
in Fig. \ref{Fig1}(a)), where he assumed the atomic volume to have
dimensions small compared with the radiation wavelength.

In our present work, we are interested in the time evolution of a specially
prepared state obtained by absorption of a single photon \cite%
{Scul06a,Scul06b,kurizki}. We report novel dynamical oscillations in the
evolution of the quantum state of the atom cloud, even without the existence
of a cavity. Fig. \ref{Fig1} summarizes the main results of this paper. It
is as if the atomic cloud acts to form a new \textquotedblleft cavity" with
the atom cloud volume $V$ replacing the virtual photon volume $\mathcal{V}_{%
\text{ph}}$ defined by the electromagnetic cavity; that is the usual vacuum
Rabi frequency $\Omega _{\text{Vac}}=(\wp /\hbar )\sqrt{\hbar \omega
/(\epsilon _{0}\mathcal{V}_{\text{ph}})}$ is replaced by $\Omega _{0}=(\wp
/\hbar )\sqrt{\hbar \omega /(\epsilon _{0}V)}$ in the present problem, where 
$\wp $ is the electric-dipole transition matrix element, $\hbar \omega $ is
the photon energy and $\epsilon _{0}$ is free space permittivity.

Here we analytically solve the equation of motion for the system in two
regimes. In one regime, we find that, for finite atom cloud size, the atomic
excitation will decay with a rate determined by the photon escape time,
together with fast dynamical oscillations (as indicated in Fig. \ref{Fig1}%
(c)) with the effective Rabi oscillation frequency $\Omega $ proportional to 
$\sqrt{N}$.

However, if the oscillation period is much greater then the time of photon
flight through the cloud, the atomic state will decay exponentially. For a
small cloud ($R\ll \lambda $) the decay rate is $N\gamma $, where $\gamma $
is the single atom decay rate, see Fig. 1a. For a larger cloud ($R\gg
\lambda $ but $\Omega R/c\ll 1$) the decay rate goes as $N\gamma (\lambda
/R)^{2}$ as in Fig. 1b. Finally, for an even larger cloud ($\Omega R/c\gg 1$%
) the probability of photon emission oscillates while decaying with a rate $%
\sim c/R$, see Fig. \ref{Fig1}c.

These two regimes (b and c) are determined by whether the system persists
memory effect, i.e., non-Markovian or Markovian regimes. We discuss
connection with previous work \cite%
{elberly2006,Cumm86,benivegna,buzek,buzek2,eberly} at the end of this paper.

We consider a system of N two-level ($a$ excited and $b$ ground) atoms,
initially one of them is in the excited state $a$ (with no information which
one), $E_{a}-E_{b}=\hslash \omega $, and the multi-mode radiation field is
in the vacuum. Atoms are located at positions $\mathbf{r}_{j}$ ($j=1,...,N$%
). The whole set of states can be expressed as those in Fig. 2 \cite{Scul06b}%
, where $|j\rangle
\,\,\,=|b_{1},b_{2},...,b_{j-1},a_{j},b_{j+1}...b_{N}\rangle $ represents
the state in which the $j^{th}$ atom is exited but the others are in the
ground state and $|g\rangle =|b_{1},b_{2},...,b_{N}\rangle $ is the state
with all the atoms in the ground state. The atomic state prepared by
uniformly absorbing one single photon with wavevector $\mathbf{k}_{0}$ is
exactly the $|+\rangle _{\mathbf{k}_{0}}$ state of Fig. 2b. In the limit $%
R\gg \lambda $ we focus here the $|+\rangle _{\mathbf{k}_{0}}$ state is
approximately an eigenstate of the system \cite{Scul06b}. This makes the
basis set of Fig. 2b preferable. The state vector for the atom-field system
at time $t$ can be then written as 
\begin{equation}
|\Psi (t)\rangle =[\beta _{+}(t)|+\rangle _{\mathbf{k}_{0}}+\beta
_{1}(t)|1\rangle _{\mathbf{k}_{0}}+\beta _{2}(t)|2\rangle _{\mathbf{k}%
_{0}}+...  \notag
\end{equation}%
\begin{equation}
+\beta _{N-1}(t)|N-1\rangle _{\mathbf{k}_{0}}]|0\rangle +\sum_{\mathbf{k}%
}\gamma _{\mathbf{k}}(t)|g\rangle |1_{\mathbf{k}}\rangle \,,
\end{equation}%
with initial conditions $\beta _{+}(0)=1$ and all other probability
amplitudes are zero. In the dipole approximation the atom-field interaction
is described by the Hamiltonian%
\begin{equation}
\hat{H}_{\text{int}}=\sum_{\mathbf{k}}\sum_{j=1}^{N}\hbar g_{k}\left[ \hat{%
\sigma}_{j}\hat{a}_{\mathbf{k}}^{\dag }\exp (i(\omega _{k}-\omega )t-i%
\mathbf{k\cdot r}_{j})+\text{adj}\right] ,  \label{d1}
\end{equation}%
\begin{widetext}

\begin{figure}[h]
  \hfill
 \framebox{ \begin{minipage}{.475\textwidth}
      \begin{center}  
      \epsfig{file=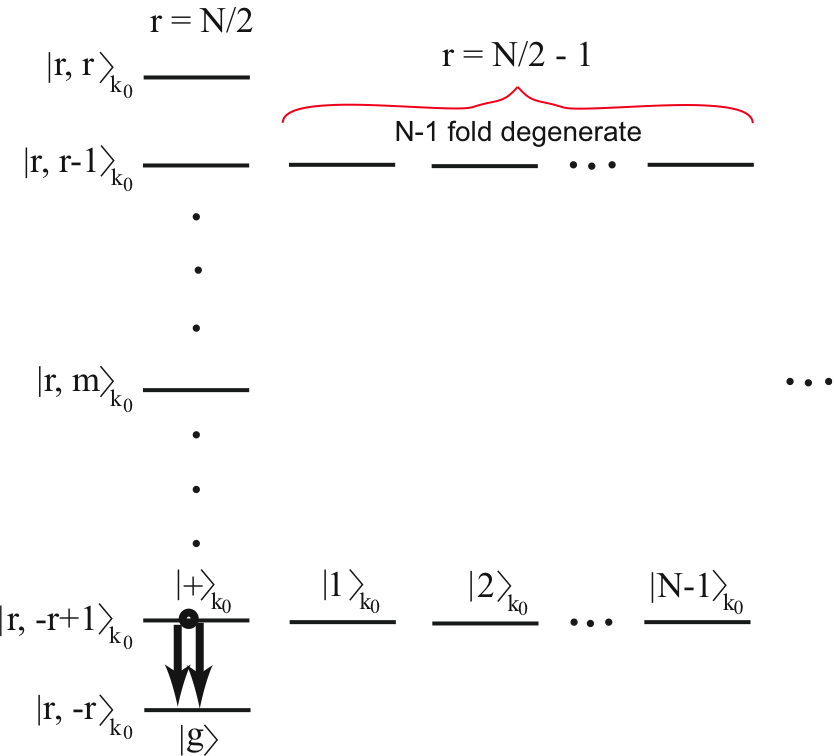, scale=0.461}
      \label{fig-tc}\\
(a)
    \end{center}
  \end{minipage}}
  \hfill
\framebox{  \begin{minipage}{.475\textwidth}
    \begin{eqnarray*}
|+\rangle _{{\mathbf{k}}_{0}} &=&\frac{1}{\sqrt{N}}\sum_{j}e^{i{\mathbf{k}}_{0}\cdot 
{\mathbf{r}}_{j}}|j\rangle \\
|1\rangle _{{\mathbf{k}}_{0}} &=&\frac{1}{\sqrt{2}}\left[ |1\rangle e^{i{%
\mathbf{k}}_{0}\cdot {\mathbf{r}}_{1}}-|2\rangle e^{i{\mathbf{k}}_{0}\cdot {%
\mathbf{r}}_{2}}\right] \\
|2\rangle _{{\mathbf{k}}_{0}} &=&\frac{1}{\sqrt{6}}\left[ |1\rangle e^{i{%
\mathbf{k}}_{0}\cdot {\mathbf{r}}_{1}}+|2\rangle e^{i{\mathbf{k}}_{0}\cdot {%
\mathbf{r}}_{2}}-2|3\rangle e^{i{\mathbf{k}}_{0}\cdot {\mathbf{r}}_{3}}%
\right] \\
&& \\
&\cdot & \\
&\cdot & \\
&\cdot & \\
|N-1\rangle _{{\mathbf{k}}_{0}} &=&\frac{1}{\sqrt{(N-1)N}}[|1\rangle e^{i{%
\mathbf{k}}_{0}\cdot {\mathbf{r}}_{1}}+|2\rangle e^{i{\mathbf{k}}_{0}\cdot {%
\mathbf{r}}_{2}}+\cdot \cdot \cdot \\
&&+|N-1\rangle e^{i{\mathbf{k}}_{0}\cdot {\mathbf{r}}_{N-1}}-(N-1)|N\rangle e^{i{%
\mathbf{k}}_{0}\cdot {\mathbf{r}}_{N}}]
\end{eqnarray*}%
 \begin{center}
 (b)
 \end{center}
  \end{minipage}}
  \hfill
  \caption{Timed Dicke states associated with absorption of radiation of wave 
vector ${\mathbf k}_{0}$: (a) the initial state $|+\rangle_{\mathbf{k}_{0}}$ 
decays directly to the graund state $|g\rangle$. (b) the timed Dicke states 
corresponding to single photon excitations. }
\end{figure}
\end{widetext}\twocolumngrid\noindent where $\hat{\sigma}_{j}$ is the
lowering operator for atom $j$, $\hat{a}_{\mathbf{k}}$ is the photon
operator and $g_{k}=\wp \sqrt{\hbar \omega _{k}/(\epsilon _{0}\mathcal{V}_{%
\text{ph}})}$ is the atom-photon coupling constant for the ${\mathbf{k}}$
mode, $\omega _{k}$ is the photon frequency and $\omega =ck_{0}$ is the
energy difference between level $a$ and $b$, $c$ is the speed of light. For
simplicity, we neglect the effects of photon polarization. The dynamical
evolution is then totally determined by the Schr\"{o}dinger's equation. Let
us consider first the two atoms problem, and call the state $|1\rangle _{%
\mathbf{k}_{0}}=|-\rangle _{\mathbf{k}_{0}}$. The state vector for the two
atoms plus field system is then given by 
\begin{equation}
|\Psi (t)\rangle =\beta _{+}(t)|+\rangle _{{\mathbf{k}}_{0}}|0\rangle +\beta
_{-}(t)|-\rangle _{{\mathbf{k}}_{0}}|0\rangle +\sum_{{\mathbf{k}}}\gamma _{{%
\mathbf{k}}}(t)|g\rangle |1_{{\mathbf{k}}}\rangle .
\end{equation}%
As is shown in Ref. \cite{Scul06b}, the probability amplitude $\beta _{+}$
and $\beta _{-}$ are coupled due to the fact that they decay to a common
ground state, that is 
\begin{equation}
\dot{\beta}_{+}=-\gamma _{++}\beta _{+}-\gamma _{+-}\beta _{-},
\end{equation}%
\begin{equation}
\dot{\beta}_{-}=-\gamma _{--}\beta _{-}-\gamma _{-+}\beta _{+}.
\end{equation}%
This was first pointed out by Agarwal \cite{agarwal}. It is closely related
to earlier work of Fano, and is often referred to as Fano coupling. We call
it Agarwal-Fano coupling.

However, when we consider a sphere with $N$ atoms, where $N\gg 1$, and $R\gg
\lambda $ there is no Agarwal-Fano coupling. Instead we now find \cite%
{Scul06b}%
\begin{equation}
\dot{\beta}_{+}(t)=-\frac{1}{N}\int_{0}^{t}dt^{\prime }\sum_{{\mathbf{k}}%
}\sum_{i,j=1}^{N}g_{k}^{2}\exp [i(\omega _{k}-\omega )(t^{\prime }-t)] 
\notag
\end{equation}%
\begin{equation}
\cdot \exp [i({\mathbf{k}}-{\mathbf{k}}_{0})\cdot ({\mathbf{r}}_{i}-{\mathbf{%
r}}_{j})]\beta _{+}(t^{\prime }).  \label{i2}
\end{equation}

For a dense cloud one can treat the atom distribution as continuous, we then
have $\sum_{i,j}\rightarrow \left( N/V\right) ^{2}\int d{\mathbf{r}}\int d{%
\mathbf{r}}^{\prime }$, where $V=4\pi R^{3}/3$ is the volume of the
spherical atomic cloud. The summation over ${\mathbf{k}}$ can also be
replaced by integration $\sum_{{\mathbf{k}}}\rightarrow \displaystyle%
\mathcal{V}_{\text{p}h}/(2\pi )^{3}\int d{\mathbf{k}},$ where $\mathcal{V}_{%
\text{ph}}$ is the photon volume. Then the equation of motion reads%
\begin{equation}
\dot{\beta}_{+}(t)=-\frac{\mathcal{V}_{\text{ph}}}{(2\pi )^{3}}\frac{N}{V^{2}%
}\int d\mathbf{k}\int d\mathbf{r}\int d\mathbf{r}^{\prime
}g_{k}^{2}\int_{0}^{t}dt^{\prime }\beta _{+}(t^{\prime })  \notag
\end{equation}%
\begin{equation}
\cdot \exp [i(\omega _{k}-\omega )(t^{\prime }-t)+i(\mathbf{k}-\mathbf{k}%
_{0})\cdot (\mathbf{r}-\mathbf{r}^{\prime })]\,.  \label{a1}
\end{equation}%
In the limit $R\rightarrow \infty $ integration over $\mathbf{r}^{\prime }$
gives the delta-function%
\begin{equation*}
\int d\mathbf{r}^{\prime }\exp [-i(\mathbf{k-k}_{0})\mathbf{r}^{\prime
}]=(2\pi )^{3}\delta (\mathbf{k-k}_{0}),
\end{equation*}%
and thus we obtain%
\begin{equation}
\dot{\beta}_{+}(t)=-N\Omega _{0}^{2}\int_{0}^{t}dt^{\prime }\beta
_{+}(t^{\prime })\,,  \label{c3}
\end{equation}%
where we have defined $\Omega _{0}=(\wp /\hbar )\sqrt{\hbar \omega
/(\epsilon _{0}V)}$, which is like vacuum Rabi frequency but with atomic
volume $V$ replacing photon volume $\mathcal{V}_{\text{ph}}$.

Differentiating both sides of Eq. (\ref{c3}) yields a harmonic oscillator
equation%
\begin{equation}
\ddot{\beta}_{+}(t)+\Omega ^{2}\beta _{+}(t)=0,  \label{c4}
\end{equation}%
where $\Omega =\sqrt{N}\Omega _{0}$ is an effective Rabi frequency.
Therefore in the limit $R\rightarrow \infty $ the atomic state undergoes
harmonic oscillations with the effective Rabi frequency $\Omega $ 
\begin{equation}
\beta _{+}(t)=\cos (\Omega t).  \label{a3}
\end{equation}

To find a solution of Eq. (\ref{a1}) at finite $R$, but yet $k_{0}R\gg 1$,
we rewrite it as%
\begin{equation}
\dot{\beta}_{+}(t)=-\frac{2\mathcal{V}_{\text{ph}}N}{\pi V^{2}}\int d\mathbf{%
k}g_{k}^{2}\int_{0}^{t}dt^{\prime }\beta _{+}(t^{\prime })e^{i(\omega
_{k}-\omega )(t^{\prime }-t)}S({\mathbf{k},R})^{2},  \label{a7}
\end{equation}%
where%
\begin{equation*}
S({\mathbf{k},R})=\frac{1}{4\pi }\int_{V}d\mathbf{r}\exp [i(\mathbf{k-k}_{0})%
\mathbf{r}]
\end{equation*}%
\begin{equation}
=\frac{\sin (|\mathbf{k-k}_{0}|R)}{|\mathbf{k-k}_{0}|^{3}}-\frac{R\cos (|%
\mathbf{k-k}_{0}|R)}{|\mathbf{k-k}_{0}|^{2}}.  \label{a6}
\end{equation}

Next we approximate $g_{k}^{2}\approx g_{k_{0}}^{2}$ and replace integration
over $\mathbf{k}$ by integration over $\mathbf{p}=\mathbf{k-k}_{0}$. The
main contribution to the integral comes from the region $p\lesssim 1/R$.
That is under the exponent one can replace $k-k_{0}\simeq \mathbf{k}%
_{0}\cdot \mathbf{p}/k_{0}$. Then Eq. (\ref{a7}) reads%
\begin{equation*}
\dot{\beta}_{+}(t)=-\frac{2}{\pi }N\Omega _{0}^{2}\int d\mathbf{p}%
\int_{0}^{t}dt^{\prime }\beta _{+}(t^{\prime })e^{[ic\mathbf{k}_{0}\cdot 
\mathbf{p}(t^{\prime }-t)/k_{0}]}
\end{equation*}%
\begin{equation}
\left[ \frac{\sin (pR)}{p^{3}}-\frac{R\cos (pR)}{p^{2}}\right] ^{2}.
\label{a8}
\end{equation}%
Integration over directions of $\mathbf{p}$ yields%
\begin{equation*}
\dot{\beta}_{+}(t)=-\frac{8}{c}N\Omega _{0}^{2}\int_{0}^{\infty
}pdp\int_{0}^{t}dt^{\prime }\beta _{+}(t^{\prime })
\end{equation*}%
\begin{equation}
\cdot \frac{\sin [cp(t^{\prime }-t)]}{(t^{\prime }-t)}\left[ \frac{\sin (pR)%
}{p^{3}}-\frac{R\cos (pR)}{p^{2}}\right] ^{2}.  \label{a9}
\end{equation}%
To integrate over $p$ we use the following formula%
\begin{equation*}
\int_{0}^{\infty }pdp\frac{\sin [cp(t^{\prime }-t)]}{(t^{\prime }-t)}\left[ 
\frac{\sin (pR)}{p^{3}}-\frac{R\cos (pR)}{p^{2}}\right] ^{2}=
\end{equation*}%
\begin{equation}
\left\{ 
\begin{array}{c}
\frac{\pi c}{96}\left[ 16R^{3}+12cR^{2}(t^{\prime }-t)-c^{3}(t^{\prime
}-t)^{3}\right] ,\quad c|t^{\prime }-t|<2R \\ 
0,\quad \text{otherwise}%
\end{array}%
\right.
\end{equation}%
which gives%
\begin{equation*}
\dot{\beta}_{+}(t)=-\frac{1}{4}N\Omega _{0}^{2}\int_{0}^{t}dt^{\prime }\beta
_{+}(t^{\prime })\left[ 16+12\frac{c}{R}(t^{\prime }-t)\right.
\end{equation*}%
\begin{equation}
\left. -\frac{c^{3}}{R^{3}}(t^{\prime }-t)^{3}\right] \Theta \left[
c(t^{\prime }-t)+2R\right] .  \label{a10}
\end{equation}%
Next we note that the function $16+12\frac{c}{R}(t^{\prime }-t)-\frac{c^{3}}{%
R^{3}}(t^{\prime }-t)^{3}$ and its derivative over $t^{\prime }$ is equal to
zero when $c(t^{\prime }-t)+2R=0$. Taking derivative of both sides of Eq. (%
\ref{a10}) twice we obtain:%
\begin{equation*}
\dddot{\beta}_{+}(t)=-N\Omega _{0}^{2}\left\{ \dot{\beta}_{+}(t)-\frac{3c}{4R%
}\beta _{+}(t)\right. -
\end{equation*}%
\begin{equation}
\left. \frac{3c^{3}}{8R^{3}}\int_{0}^{t}dt^{\prime }\beta (t^{\prime
})(t^{\prime }-t)\Theta \left[ c(t^{\prime }-t)+2R\right] \right\} .
\label{a12}
\end{equation}%
Next we assume that 
\begin{equation}
N\Omega _{0}^{2}\frac{R^{2}}{c^{2}}\gg 1,\quad \text{or}\quad \frac{\Omega R%
}{c}\gg 1.  \label{a13}
\end{equation}%
Then one can omit the last term in Eq. (\ref{a12}) which yields%
\begin{equation}
\dddot{\beta}_{+}(t)+\Omega ^{2}\dot{\beta}_{+}(t)-\frac{3c\Omega ^{2}}{4R}%
\beta _{+}(t)=0.  \label{a14}
\end{equation}%
Solution of Eq. (\ref{a14}) under the condition (\ref{a13}) is given by%
\begin{equation}
\beta _{+}(t)=\cos (\Omega t)\exp \left( -\frac{3c}{8R}t\right) ,
\label{a16}
\end{equation}%
which describes rapid oscillations with the effective Rabi frequency $\Omega 
$ superimposed by the exponential decay. The state decays during the time of
the photon flight through the atomic cloud. The emitted photon is reabsorbed
and reemitted many times before it leaves the cloud.

In the opposite limit, $\Omega R/c\ll 1$, one can use the Markovian
approximation. We integrate Eq. (\ref{a10}) over $t^{\prime }$ assuming $%
\beta _{+}(t^{\prime })$ is a slow varying function of $t^{\prime }$ and
approximate $\beta _{+}(t^{\prime })\approx \beta _{+}(t)$. Then for $t>2R/c$
we obtain%
\begin{equation}
\dot{\beta}_{+}(t)=-\Gamma \beta _{+}(t),  \label{c51}
\end{equation}%
which yields an exponentially decaying solution%
\begin{equation}
\beta _{+}(t)=\beta _{+}(0)e^{-\Gamma t}.
\end{equation}%
Here $\Gamma =3\Omega ^{2}R/4c=27N\gamma /8(k_{0}R)^{2}$ and $\gamma =%
(\omega ^{3}\wp ^{2})/(6\pi \epsilon _{0}\hbar c^{3})$ is the spontaneous
decay rate for one atom.

\textit{Discussion:} Similar problems have been investigated in the past
several decades. Cummings \cite{Cumm86} considered the spontaneous emission
of a single atom which is initially excited in the presence of $N-1$
initially unexcited identical atoms, when there are $M$ accessible radiation
modes. He showed that such an extended system oscillates between the ground
sate and the excited state with an effective Rabi frequency $\Omega \sim 
\sqrt{N}$. Such modification of the spontaneous emission of one atom in the
presence of $N-1$ atoms inside a cavity has been studied since then \cite%
{benivegna}, \cite{buzek}. Buzek \cite{buzek2} studied the dynamics of an
excited atom in the presence of $N-1$ unexcited atoms in the free space and
predicted that there is a radiation suppression but did not report dynamical
oscillations.

The effective Rabi frequency $\Omega =\sqrt{N}\Omega _{0}$ we\ found from
quantum mechanical consideration can be written as $\Omega =\sqrt{(3/4\pi
^{2})\gamma \omega (N/V)\lambda ^{3}}=\sqrt{n\omega \wp ^{2}/\epsilon
_{0}\hbar }$. This result is analogous to the plasma frequency and can be
obtained in a classical model by treating atoms as classical harmonic
oscillators \cite{burnham}. Indeed, replacing the electric-dipole transition
matrix element by $\wp =e\cdot d$, where $d=\sqrt{\hslash /m\omega }$ is the
oscillator length, yields precisely the plasma frequency $\Omega =\sqrt{%
ne^{2}/m\epsilon _{0}}$.

Relevant experiments have been carried out by the groups of Lukin \cite%
{lukin}, Kuzmich \cite{kuzmich}, Kimble \cite{kimble}, Vuleti$\acute{\text{c}%
}$ \cite{vuletic}, Harris \cite{Bali05} et al. \cite{note}. For realistic
physical situations such as $n=10^{14}$cm$^{-3}$, $\omega /2\pi =6\times
10^{14}$Hz, $R=10$ cm ($N=4\times 10^{17}$) and $|\wp |=10^{-29}$C$\cdot $m,
we obtain that the state decay is accompanied by a few oscillations with the
effective Rabi frequency $\Omega \approx 2\times 10^{10}$Hz and the decay
time is about $R/c\sim 3\times 10^{-10}$s. One can observe a crossover to
the exponentially decaying regime, e.g., by decreasing the size of the
atomic cloud.

In summary, we study correlated spontaneous emission of a totally symmetric $%
N$-atom state prepared by an absorption of a single photon. This is an
extension of the result obtained in Refs. \cite{Scul06a,Scul06b}. Decay of
such a state occurs via photon emission in the direction of the incident
photon for large enough density. We found that time evolution of the initial
state depends on the relation between an effective Rabi frequency $\Omega
\propto \sqrt{N}g_{k_{0}}$ and the time of photon flight through the cloud $%
R/c$. If $\Omega R/c<1$ the state exponentially decays with the rate $\Omega
^{2}R/c$ which is determined by the Dicke superradiance rate $N\gamma $
reduced by the factor of $\lambda ^{2}/R^{2}$ due to smaller finite state
phase volume in the case of directional emission. In the opposite limit $%
\Omega R/c\gg 1$ the decay is accompanied by oscillations with the effective
Rabi frequency $\Omega $ and the decay time is given by $R/c$.

We gratefully acknowledge the support of the Office of Naval Research (Award
No. N00014-03-1-0385) and the Robert A. Welch Foundation (Grant No. A-1261).


\begin{thebibliography}{99}
\bibitem{note} To put the present results in context we emphasize the
difference between the physics of the groups of Kimble, Harris and Lukin and
the present work. In their experiment the Rabi oscillations in the two
photon correlation function are governed by the probe laser strength and
decay with the atomic decay rate. In the present work $\Omega \sim \sqrt{N/V}
$ and the state decays with a rate determined by the time of photon flight
across the atomic cloud.
\end{thebibliography}
\end{document}